\documentclass[showpacs,preprintnumbers,amsmath,amssymb]{revtex4}
\usepackage{graphicx}% Include figure files
\usepackage{dcolumn}% Align table columns on decimal point
\usepackage{bm}% bold math

\def\mapgeq{\mathbin{\lower.3ex\hbox{$\buildrel>\over{\smash{\scriptstyle\sim}\vphantom{_x}}$}}}
\def\mapleq{\mathbin{\lower.3ex\hbox{$\buildrel<\over{\smash{\scriptstyle\sim}\vphantom{_x}}$}}}
\def\mapgeqeq{\mathbi{\lower.3ex\hbox{$\buildrel>\over{\smash{\scriptstyle\approx}\vphantom{_2}}$}}}
\def\mapleqeq{\mathbin{\lower.3ex\hbox{$\buildrel<\over{\smash{\scriptstyle\approx}\vphantom{_2}}$}}}
\def\Journal#1#2#3#4{{#1} {\bf #2} (#4) #3}
\def\NPB{Nucl. Phys. B}
\def\NPBOLD{Nucl. Phys.}
\def\NPSUPPL{Nucl. Phys. Proc. Suppl.}
\def\PLB{{Phys. Lett.} B}

\def\PLBOLD{Phys. Lett.}
\def\PRL{Phys. Rev. Lett.}
\def\RMP{Rev. Mod. Phys.}
\def\PRD{Phys. Rev. D}

\def\PTPSUPPL{Prog. Theor. Phys. Suppl.}
\def\PTP{Prog. Theor. Phys.}

\def\PPNP{Prog. Part. Nucl. Phys.}

\begin{document}

\preprint{TOKAI-HEP/TH-0301}

\title{
Flavor Symmetry Breaking in Strongly Coupled $N$=1\\Supersymmetric $SO(N_c)$ Gauge Theory with $N_f=N_c-2$~\footnote{to be published in Europhysics Letters (2003)}}% Force line breaks with \\

\author{Masaki Yasu\`{e}}
% \homepage{}
\email{yasue@keyaki.cc.u-tokai.ac.jp}
\affiliation{%
\sl Department of Physics, Tokai University,\\
1117 Kitakaname, Hiratsuka,\\
Kanagawa 259-1291, Japan.}%

\date{April, 2003}% It is always \today, today,
             %  but any date may be explicitly specified

%%-------------------------------------------------
%% Abstruct
%%-------------------------------------------------
\begin{abstract}
In the $SO(N_c)$ gauge theory with $N_f$ quarks for $N_f=N_c-2$, its instanton effects indicate the signal of dynamical flavor symmetry breakdown of $SU(N_f)$ to $SO(N_f)$, which is not described by the conventional ``magnetic" degree's of freedom.  It is argued that this breaking is well described by our effective superpotential consisting of ``electric" quarks and gluons instead of monopoles of $SO(N_c)$.  The low-energy particles include the Nambu-Goldstone superfields associated with this breakdown.  The proposed superpotential is found to exhibit the holomorphic decoupling property and the anomaly-matching property on a residual chiral $U(1)$ symmetry.
\end{abstract}

\pacs{PACS: 11.15.Ex, 11.15.Tk, 11.30.Rd, 11.30.Pb}
% PACS, the Physics and Astronomy
                             % Classification Scheme.
%\keywords{Keywords: neutrino mass, radiative mechanism, lepton triplet}%Use showkeys class option if keyword
                              %display desired
\maketitle

%%-------------------------------------------------
%% Main body
%%-------------------------------------------------
%%-------------------------------------------------
%% Main body
%%-------------------------------------------------
%%%%%%%%%%%%%%%%%%%%%%%%%%%%%%%%%%%%%%%%%%%%%%%%%%%%%%%%%%%%%%%%%%%%%%%%%%%%%%%%%%%%%%%%
%\section{\bf Introduction}
%\section{\bf Introduction}
It is well known that the physics of strongly coupled $N$=1 supersymmetric (SUSY) gauge theories can be described by the $N=1$ duality using the ``magnetic" degrees of freedom instead of the ``electric" degrees of freedom \cite{Seiberg}.  This description has been motivated by the great success of the $N=2$ duality in $N=2$ SUSY theories, where the ``magnetic" degrees of freedom determine various properties of $N=2$ physics \cite{SeibergWitten}. However, neither the appearance of the ``magnetic" degrees of freedom can be readily understood inside $N=1$ SUSY theories nor the meaning of the $N=1$ duality is clear. The physics due to the $N=1$ duality can be traced back to the corresponding physics due to the $N$=2 duality \cite{BrokenN_2}.  Especially, the ``electric" quarks and gluons in $N=2$ SUSY theories are continuously deformed to the ``magnetic" quarks and gluons in the corresponding $N=1$ SUSY theories.  It is also known that $N=2$ SUSY theories are too constrained to cover physics realized in all of $N=1$ SUSY theories.  For example, supersymmetric quantum chromodynamics (SQCD) with $N_c$-colors and $N_f$-flavors exhibits the $N$=1 duality \cite{Seiberg,EarlySeiberg} only for $3N_c/2< N_f$ ($<$ 3$N_c$), where its phase is characterized as an interacting Coulomb phase \cite{CoulombPhase}.  It is, thus, obscure that SQCD with ($N_c+2\leq$) $N_f\leq 3N_c/2$ really enjoys the $N=1$ duality although it satisfies plenty of indirect consistency conditions \cite{Review} including those from brane constructions in string theory \cite{Brane}.  The similar questioning on the validity of the $N=1$ duality has lately appeared in $SO(N_c)$ SUSY theories (with $N_f< 3(N_c-2)$) \cite{SO_IS,SO} by demonstrating that physics near the Chebyshev point (known as a superconformal point \cite{Conformal}) in explicitky broken $N=2$ SUSY theories is found to be characterized by the presence of dynamical flavor symmetry breaking \cite{RecentSO}, which should be contrasted with the absence of flavor symmetry breaking in the $N$=1 duality.  It is still possible to expect that the Chebyshev point may merge with the ordinary singular point in the exact $N=1$ limit.

As an alternative description of the physics of strongly coupled $N$=1 SUSY theories \cite{Alternative,YasueSU,YasueSO}, which is not directly supported by the $N=2$ duality, we have suggested to use Nambu-Goldstone superfields as low-energy degrees of freedom associated with dynamically flavor symmetry breaking \cite{YasueSU,YasueSO}. Namely, the suggested phase is described by confined ``electric" quarks and gluons instead of ``magnetic" quarks and gluons. The essence to make dynamical breakdown of flavor symmetry more visible in N=1 SUSY theories lies in the inclusion of a composite superfield ($S$) made of two chiral gauge superfields as well as mesons and baryons \cite{VY}. This composite superfield has been excluded in the low-energy degrees of freedom because one can envision that it turns out heavy after the vacuum structure is determined.  But, it will correctly describe the instanton physics present in $N=1$ SUSY theories and has been proved that our superpotentials constructed along this line of reasoning in fact reproduce the consistent vacuum with the instanton effects, where $N_c$ (or $N_c-2$) quarks get heavy while others remain massless in the $SU(N_c)$ (or $SO(N_c)$) SUSY theories with $N_f\geq N_c+2$ (or $N_f\geq N_c$) \cite{YasueSU,YasueSO}.  Furthermore, it has greatly helped us reveal the existence of the vacuum located away from the origin of a moduli space.  Of course, this new phase coexists with the one with the $N=1$ duality as much as the same way that QCD with two flavors theoretically allows both phases with massless proton and neutron (corresponding to the ``magnetic" description) as well as with pions (corresponding to the ``electric" description) to exist.  

One may raise the objection stating that this kind of dynamical flavor symmetry breaking cannot occur in strong coupling SUSY gauge theories since the $N$=1 duality description is valid and the emergence of the origin of moduli space forbids the presence of dynamical symmetry breaking.  In this paper, we would like to show affirmative discussions on the presence of the flavor symmetry breaking, thus located away from the origin of moduli space, in the $SO(N_c) $ theory with $N_f=N_c-2$ along the above line of thought, which should be contrasted with the absence of the flavor symmetry breaking based on the $N$=1 duality, utilizing two monopoles \cite{SO}.  We rely upon the known instanton dynamics of the $SO(N_c)$ theory with $N_f=N_c-2$, leading to the flavor symmetry breaking of $SU(N_f)$. However, so far no superpotential has been proposed to reproduce this instanton effect in the ``electric" phase.  We describe this dynamics in the form of an effective superpotential for $N_f=N_c-2$ regulated by ``electric" degrees of freedom. 

In the $SO(N_c)$ theory with $N_f$-quarks, the flavor symmetry is given by
%%%%%%%
\begin{equation}\label{Eq:FlavorSym}
 G = SU(N_f)\times U(1)_R,
\end{equation}
%%%%%%%
under which chiral quark superfields of $Q^i_A$ and chiral gauge superfields of $W_{[AB]}$
%%%%%%%
 \footnote{Throughout this paper, the labels of $i,j,\cdots$ specify the flavor indices of $SU(N_f)$ and those of $A,B,\cdots$ specify the color indices of $SO(N_c)$}  
%%%%%%%
transform according to TABLE \ref{Tab:QuntumNumber}.  The instanton of $SO(N_c)$ yields the contributions from the gluinos and $N_f$ quarks, which can be expressed as 
%%%%%%%
%\begin{equation}
%\label{Eq:Instanton}
$(\lambda\lambda)^{N_c-2}\det (\psi^i\psi^j)$, 
%\end{equation}
%%%%%%%
where $\psi$ ($\lambda$) is a spinor component of $Q$ ($S$).  This instanton amplitude is equivalent to
%%%%%%%
\begin{equation}
\label{Eq:InstantonScalar}
\det (\phi^i\phi^j) \propto \prod_{i=1}^{N_f}\pi_i \sim \Lambda^{2N_f},
\end{equation}
%%%%%%%
where $\phi^i$ is the $i$-th scalar quark and $\pi_i$ represents the scalar component of $T^{ii}$.  This evaluation clearly shows the appearance of the dynamical flavor symmetry breaking of
%%%%%%%
\begin{equation}
\label{Eq:FlavorSymBreaking}
SU(N_f) \times U(1)_R \rightarrow SO(N_f) \times U(1)_R,
\end{equation}
%%%%%%%
with $SO(N_f)$ as a maximal subgroup. 
%%%%%%%%%%%%%%%
\begin{table}[!t]
    \begin{center}
    \begin{tabular}{|c||c|c|c|}
    \hline
        field  & $SO(N_c)$ & $SU(N_f)$ & $U(1)_R$\\ \hline
         $Q^i_A$  & ${\bf N_c}$ & ${\bf N_f}$ & $(N_f-N_c+2)/N_f$ \\
         $W_{[AB]}$ & {\bf ADJ} & {\bf 1} & 1 \\\hline
    \end{tabular}
    \end{center}
    \caption{\small \label{Tab:QuntumNumber} Quantum numbers of quarks, $Q^i_A$, and gauge fields, $W_{[AB]}$, where {\bf ADJ} stands for the adjoint representation of $SO(N_c)$.}
\end{table}
%%%%%%%%%%%%%%%

We would like to explain this instanton physics realized in the ``electric" phase of $SO(N_c)$ on the basis of an effective superpotential. Possible composite superfields involved in our discussions are chiral superfields (mesons) of
%%%%%%%
%\begin{equation}\label{Eq:FieldContentT}
% T^{ij}  =  \sum_{A=1}^{N_c} Q_A^iQ_A^j,  
$ T^{ij}  =  \Sigma_{A=1}^{N_c} Q_A^iQ_A^j$,  
%\end{equation}
%%%%%%%
and a chiral flavor-singlet ``gauge" superfield (baryons) \cite{SO_IS} of
%%%%%%%
%\begin{eqnarray}\label{Eq:FieldContentB}
% B &=& 
%\sum_{i_1\dots i_{N_c-2}}
%\sum_{A_1\dots A_{N_c}}
%\frac{1}{(N_c-2)!(N_c-2)!2!}
%\varepsilon_{i_1i_2{\dots}i_{N_c-2}}
%\varepsilon^{A_1A_2{\dots}A_{N_c}}
%\nonumber \\
%&& Q_{A_1}^{i_1}\dots Q_{A_{N_c-2}}^{i_{N_c-2}}W_{[A_{N_c-1}A_{N_c}]},
%\end{eqnarray}
%%%%%%%
$B$ = $\Sigma_{i_1\dots i_{N_c-2}}\Sigma_{A_1\dots A_{N_c}}$
 $\varepsilon_{i_1i_2{\dots}i_{N_c-2}}$ $\varepsilon^{A_1A_2{\dots}A_{N_c}}$
$Q_{A_1}^{i_1}\dots Q_{A_{N_c-2}}^{i_{N_c-2}}$ $W_{[A_{N_c-1}A_{N_c}]}$ / 
$(N_c-2)!(N_c-2)!2!$
 as well as a composite superfield made of two chiral gauge superfields
%%%%%%%
%\begin{equation}\label{Eq:FieldContentS}
%S = \frac{g^2}{32\pi^2}\sum_{A,B=1}^{N_c} W_{[AB]}W_{[AB]},
%\end{equation}
%%%%%%%
$S$ = $g^2\Sigma_{A,B=1}^{N_c} W_{[AB]}W_{[AB]}/32\pi^2$ 
with the $SO(N_c)$ gauge coupling, $g$, where the gauge coupling is explicitly included in $S$ for later discussions.  The inclusion of $S$ is advocated by Veneziano and Yankielowicz \cite{VY} some times ago while the formulation without $S$ has been motivated by Afleck, Dine and Seiberg \cite{EarlySeiberg}.  The role of $S$ is to reproduce the correct amount of the breaking of an anomalous $U(1)$ symmetry by instantons, which may not be physically required since this symmetry is not a conserved symmetry so that effective interactions need not respect its presence.  However, we adopt the description in terms of $S$ to evaluate the instanton effects by the effective superpotential approach.

The classic construction \cite{VY,MPRV,YYY} of effective superpotentials requires that not only they are invariant under the transformations of all the symmetries but also they are compatible with the response from an anomalous $U(1)_{anom}$ symmetry, namely,  $\delta{\cal L}$ $\sim$ $F^{\mu\nu}{\tilde F}_{\mu\nu}$, where ${\cal L}$ represents the lagrangian of the $SO(N_c)$ theory and $F^{\mu\nu}$ (${\tilde F}_{\mu\nu}\sim \epsilon_{\mu\nu\rho\sigma}F^{\rho\sigma}$) is a gauge field strength. Its SUSY version applied to the effective superpotential denoted by $W_{\rm eff}$ becomes
%%%%%%%
%\begin{eqnarray}\label{Eq:WeffAnomS}
$\delta W_{\rm eff} \sim  S$. 
%\end{eqnarray}
%%%%%%%
The resulting superpotential takes the simple form of
%%%%%%%
\begin{equation}\label{Eq:Weff}
W_{\rm eff}=S 
\ln\left(
\frac{
	\det T
}
{
	\Lambda^{2N_f}
}
\right),
\end{equation}
%%%%%%%
where $\Lambda$ is the mass scale of $SO(N_c)$.  We can also construct another effective superpotential including the chiral ``gauge" superfield of $B$:
%%%%%%%
\begin{equation}\label{Eq:WeffWithB}
W_{\rm eff}=S 
\ln\left(
\frac{
	\det T
}
{
	\Lambda^{2N_f}
}
\right) + f\frac{BB}{\det T},
\end{equation}
%%%%%%%
where $f$ is a coupling parameter, which also satisfies $\delta W_{\rm eff} \sim  S$
%Eq.(\ref{Eq:WeffAnomS})
.  However, this superpotential generates a massive $B$ superfield for the instanton-suggested vacuum of $\det T\neq 0$ and the anomaly-matching conditions on $U(1)_R$ fail to be satisfied.  One may worry about the classical limit of our superpotential of Eq.(\ref{Eq:Weff}), where the non-perturbative effects should vanish.  Recalling the relation of $g^{-2} \sim -[(3(N_c-2)-N_f)/(16\pi^2)]\ln(\Lambda^2/\mu^2)$, where $\mu$ is a reference scale, one can readily find that our superpotential simply yields $\propto W_{[AB]}W_{[AB]}$, which is the tree superpotential for the gauge kinetic term as stressed in Ref.\cite{VY}.

The SUSY vacuum defined by our superpotential, which requires $\partial W_{\rm eff}/\partial \pi_{i,\lambda}(\equiv W_{{\rm eff};i,\lambda})$ to vanish, is specified by
%%%%%%%
%\begin{equation}
%W_{{\rm eff};i} = \frac{\pi_\lambda}{\pi_i} = 0, \qquad
%W_{{\rm eff};\lambda}  = \ln\left(\prod\limits_{i = 1}^{N_f}\frac{\pi_i}{\Lambda^2}\right)% = 0,
%\label{Eq:Weff_iS}
%\end{equation}
%%%%%%%
$W_{{\rm eff};i}$ = $\pi_\lambda/\pi_i$ = 0 and $W_{{\rm eff};\lambda}$ = $\ln (\Pi_{i = 1}^{N_f}\pi_i/\Lambda^2)$ = 0, 
where $\pi_\lambda$ is the scalar component of $S$ \footnote{Other fields such as scalar components of $T^{ij}$ with $i\neq j$ can be set to vanish at the minimum; therefore, we omit these terms.}. The solution to these equations is given by $\Pi_{i = 1}^{N_f}\pi_i=\Lambda^{2N_f}$ and $\pi_\lambda$ = 0, leading to
%%%%%%%
\begin{equation}
\det T =\Lambda^{2N_f}, \quad S = 0.
\label{Eq:Solutions_iS}
\end{equation}
%%%%%%%
This vacuum is the same as Eq.(\ref{Eq:InstantonScalar}) generated by the instantons.  As a result, the maximal flavor symmetry respected by the $SO(N_c)$ theory turns out to be $SO(N_f) \times U(1)_R$ as in Eq.(\ref{Eq:FlavorSymBreaking}). In this end, the superpotential vanishes but the chiral Nambu-Goldstone superfields are produced as the low-energy degrees of freedom \footnote{Of course, interactions of the Nambu-Goldstone superfields arise from kinetic terms via their K$\ddot{\rm a}$hlar potentials}.  This situation is similar to the one encountered in supersymmetric quantum chromodynamics (SQCD) with $N_f=N_c$ \cite{Review,MPRV}. The corresponding superpotential in terms of $S$ is given by 
%%%%%%%
\begin{equation}\label{Eq:Weff_SQCD}
W_{\rm eff}=S 
\ln\left(
\frac{
	\det T-{\tilde B}{\bar {\tilde B}}
}
{
	\Lambda^{2N_f}
}
\right),
\end{equation}
%%%%%%%
where ${\tilde B}$ and ${\bar {\tilde B}}$ are flavor-singlet chiral superfields (baryons) analogous to $B$
%Eq.(\ref{Eq:FieldContentB})
 without $W_{[AB]}$, respectively, made of quarks and anti-quarks.  The solution is given by 
%%%%%%%
\begin{equation}
\det T-{\tilde B}{\bar {\tilde B}} = \Lambda^{2N_f}, \quad S = 0.
\label{Eq:Solutions_iS2}
\end{equation}
%%%%%%%
indicating flavor symmetry breaking in SQCD with $N_f$=$N_c$.

%%%%%%%%%%%%%%%
\begin{table}[!t]
    \begin{center}
    \begin{tabular}{|c||c|c|c|}
    \hline
        fields        & $SO(2)$     & $SO(N_f)$   & $U(1)_R$\\ \hline
         $Q^{\{ij\}}$ & {\bf 1}   & {\bf SYM}  & 0 \\
         $W^{[AB]}$   & {\bf ADJ}(={\bf 1}) & {\bf 1}   & 1 \\ \hline
         $T^{ij}$ & {\bf 1}   & {\bf SYM}  & 0 \\
         $B$          & {\bf 1}   & {\bf 1}   & 1 \\ \hline
    \end{tabular}
    \end{center}
    \caption{\label{Tab:COMP}\small Quantum numbers of superfields in the Higgs phase (the upper column) and the confining phase (the lower column), where {\bf SYM} stands for the symmetric representation of $SO(N_c)$ and $Q^{\{ij\}}$ denotes the symmetrization with respect to $i$ and $j$ in $Q^i_j$.}
\end{table}
%%%%%%%%%%%%%%%
The anomaly-matching property is easily seen by the use of the complementarity \cite{tHooft,Susskind}, where $SU(N_f) \times U(1)_R$ is also broken by $\langle 0 \vert\phi^i_A\vert 0 \rangle$ = $\delta^i_A\Lambda$, resulting in $SO(N_f) \times U(1)_R$. The low-energy degrees of freedom listed in TABLE \ref{Tab:COMP} are supplied by $Q^{\{ij\}}$ as the symmetric representation of $SO(N_f)$ and $W_{[AB]}$ as the chiral gauge superfields of $SO(2)$ arising from the leftover piece in $SO(N_c)$.  These superfields directly corresponds to the Nambu-Goldstone superfields of $T^{ij}$ (with Tr($T^{ij}$) = 0) associated with $SU(N_f)\rightarrow SO(N_f)$ and to the chiral flavor-singlet ``gauge" superfield of $B$. The field $T$ with Tr($T$)$\neq$0 forms a mass term with $S$ and the both are decoupled.  The presence of $U(1)_R$ requires that the ``gauge" superfield of $B$ to be massless.  This masslessness of the chiral ``gauge" superfield of $B$ is the indication of the appearance of a gauged $U(1)$ symmetry \cite{SO_IS}.  Since the Nambu-Goldstone superfields are neutral under this $U(1)$ symmetry, this ``photon" does not interacts with matter fields.  

This description of the $SO(N_c)$ gauge theory with $N_f=N_c-2$ should be compared with the one given by the $N=1$ duality with ``magnetic" quarks \cite{SO}.  The flavor symmetry breaking is absent and the low-energy degrees of freedom are supplied by two monopoles and the chiral ``gauge" superfield of $B$, whose anomalies match the original anomalies.  One may think that since, in the ``magnetic" phase, the flavor $SU(N_f)$ group is unbroken, the anomaly-matching on its subgroup of $SO(N_c)$ is also realized and that, furthermore, our low-energy spectrum looks the same as the one in the ``magnetic" phase. So, one may also think that it is trivial to have the consistent description of the flavor $SO(N_f)$ group.  But, our physics is regulated by the ``electric" quarks but not by the ``magnetic" quarks.  As long as our ``electric" physics also correctly describes the $SO(N_c)$ physics, it is not surprising that the both descriptions result in the similar low-energy spectrum.

The holomorphic decoupling property of Eq.(\ref{Eq:Weff}) can be seen by adding a mass of $m_{N_f}$ to the $N_f$-th quark, yielding
%%%%%%%
\begin{equation}\label{Eq:WeffMass}
W_{\rm eff}=S 
\left[ 
\ln
\left(
\frac{
S^{N_c-N_f-2}\det T
}
{
	\Lambda^{3N_c-N_f-6}
}
\right)
+N_f-N_c+2
\right]
-m_{N_f}T^{N_fN_f}.
\end{equation}
%%%%%%%
with $N_f=N_c-2$.  The vacuum is, then, specified by $S/T^{N_fN_f}=m_{N_f}$, which results in
%%%%%%%
\begin{equation}\label{Eq:WeffMass2}
W_{\rm eff}=S 
\left[ 
\ln
\left(
\frac{
S^{N_c-N_f-1}\det {\tilde T}
}
{
	{\tilde \Lambda}^{3N_c-N_f-5}
}
\right)
+N_f-N_c+1
\right]
,
\end{equation}
%%%%%%%
where $\tilde T$ is a $(N_f-1)\times(N_f-1)$ submatrix of $T$ and ${\tilde \Lambda}^{3N_c-N_f-5}$ = $m_N\Lambda^{3N_c-N_f-6}$.  This superpotential is nothing but Eq.(\ref{Eq:WeffMass}) by letting $N_f\rightarrow N_f-1$ with $T^{N_fN_f}$ decoupled. For $N_f\leq N_c-3$, after eliminating $S$ in Eq.(\ref{Eq:WeffMass}), we reach
%%%%%%%
\begin{equation}\label{Eq:ADSlike}
W^\prime_{\rm eff}=\left( N_f-N_c+2\right) \left[ 
\frac{
	\det \left(T\right)
}
{
	\Lambda^{3N_c-N_f-6}
} 
\right]^{1/(N_f-N_c+2)}, 
\end{equation}
%%%%%%%
which is the same effective superpotential as one of the series of the superpotentials discussed in Ref.\cite{SO_IS} for $N_f\leq N_c-3$.

Finally, we comment on the holomorphic decoupling from the $SO(N_c)$ theory with $N_f\geq N_c$.  We have advocated to use the effective superpotential in the ``electric" phase of $SO(N_c)$ for $N_f\geq N_c$ \cite{YasueSO}:
%%%%%%%
\begin{equation}
\label{Eq:W_eff_summary}
W_{\rm eff}=S 
\left\{ 
\ln\left[
\frac{
	S^{N_c-N_f-2}\det T\cdot f(Z)
}
{
	\Lambda^{3N_c-N_f-6}
} 
\right] 
+N_f-N_c+2\right\},
\end{equation}
%%%%%%%
where ${\hat B}$ is a baryon made of $N_c$ quarks, namely, ${\hat B} = \varepsilon^{A_1\cdots A_{N_c}}Q^{[1}_{A_1}\cdots Q^{N_c]}_{A_{N_c}}$, and $f(Z)=(1-Z)^\rho$ ($\rho > 2$) with $Z$=${\hat B}T^{N_f-N_c}{\hat B}/\det T$ \footnote{In Ref.\cite{YasueSO}, Eq.(8) should reads $f(Z)=(1-Z)^\rho$ ($\rho > 2$) and the related parts should be modified accordingly.}. The combination of ${\hat B}T^{N_f-N_c}{\hat B}$ with the obvious notation is classically equivalent to det$T$.  The spontaneous flavor symmetry breaking is also induced at the SUSY vacuum specified by this effective superpotential.  It can be found that this superpotential correctly describes the instanton effects for the $N_f-N_c+2$ of massive quarks, thus leaving $N_c-2$ quarks massless, with any finite masses and exhibits the holomorphic decoupling property for $N_f\geq N_c$ \cite{YasueSO}. 

Similarly, we also expect the holomorphic decoupling from $SO(N_c)$ for $N_f=N_c$ with the $N_f$-th quark decoupled to $SO(N_c)$ for $N_f=N_c-1$, which gives the superpotential
%%%%%%%
\begin{equation}\label{Eq:Weff3}
W_{\rm eff}=S 
\left[ 
\ln
\left(
\frac{
S^{-1}\det T\cdot f(Z)
}
{
	\Lambda^{2N_f-3}
}
\right)
+1
\right],
\end{equation}
%%%%%%%
for $N_f=N_c-1$, where $Z$ contains the field ${\hat B}^\prime$ analogous to ${\hat B}$, which can be literally obtained from ${\hat B}$ by decoupling the $N_c$-th quark: ${\hat B}^{\prime A} = \varepsilon^{AA_1\cdots A_{N_c-1}} Q^{[1}_{A_1}\cdots Q^{N_c-1]}_{A_{N_c-1}}$.  As a result, $Z$=${\hat B}{\hat B}/\det T$ for $N_f=N_c$  with the $N_c$-th quark removed is converted into $Z$=${\mathcal B}/\det T$ for $N_f=N_c-1$, where ${\mathcal B} = {\hat B}^{\prime A}{\hat B}^{\prime A}$.  Namely, a composite superfield of the ``diquark"-type given by ${\hat B}^{\prime A}$ should be present to form a color-singlet composite of ${\mathcal B}$. If $S$ is correctly decoupled, Eq.(\ref{Eq:Weff3}) becomes
%%%%%%%
\begin{equation}\label{Eq:Weff4}
W^\prime_{\rm eff}=
\frac{
\det T\cdot f(Z)
}
{
	\Lambda^{2N_f-3}
}.
\end{equation}
%%%%%%%
Since $\det T\neq 0$ (as we expected) is allowed if $f(Z)=0$, yielding $Z=1$, at the SUSY minimum, this superpotential will describe the spontaneous symmetry breaking by $Z$=1 leading to $\det T = {\mathcal B}\sim \Lambda^{2(N_c-1)}$ and $SU(N_f)\times U(1)_R\rightarrow SO(N_f)$ is realized.  Such a nonvanishing $\det T$ is found to be ensured by requiring $f^\prime (Z)=0$. For $f(Z)=(1-Z)^\rho$, it is satisfied by $\rho > 1$.  Therefore, the choice of $\rho > 2$ required for the $SO(N_c)$ theory with $N_f\geq N_c$ respects the holomorphic decoupling.  The complementarity argument follows for $Q^i_A \sim \Lambda \delta^i_A$ ($i,A=1\sim N_c-1$) and the resulting spectrum in the Higgs phase coincides with that of the Nambu-Goldstone superfields in the confining phase.  One of the linear combinations of Tr$(T)$ and ${\mathcal B}$ is decoupled by forming a mass term with $S$ and the presence of ${\mathcal B}$ is essential for the consistent decoupling of $S$. It is readily understood that the further decoupling of the $(N_c-1)$-th quark yields Eq.(\ref{Eq:Weff}).  The details can be found in the subsequent article \cite{Future}.

Summarizing our discussions, we have stressed the importance of the instanton physics specific to $SO(N_c)$ with $N_f=N_c-2$ to reveal the non-perturbative ``electric" physics of $SO(N_c)$, which indicates the spontaneous flavor symmetry breaking.  This dynamics is opposed to the $SO(N_c)$ dynamics based on the $N=1$ duality, describing the unbroken flavor symmetry realized in the ``magnetic" phase.  We have found that the following effective superpotential correctly describes the $SO(N_c)$ physics realized in the ``electric" phase with $N_f=N_c-2$:
%%%%%%%
\begin{equation}\label{Eq:Weff_1}
W_{\rm eff}=S 
\ln\left(
\frac{
	\det T
}
{
	\Lambda^{2N_f}
}
\right).
\end{equation}
%%%%%%%
Our superpotential respects
\begin{enumerate}
\item correct vacuum structure with instanton physics, 
\item holomorphic decoupling property for $N_f\leq N_c-2$ (and also for $N_f > N_c-2$ \cite{Future}),
\item dynamical breakdown of the flavor $SU(N_f)$ symmetry:
%%%%%%%
\begin{eqnarray}
\label{Eq:SymmetryBreakingSummary}
& &SU(N_f) \times U(1)_R \rightarrow SO(N_f) \times U(1)_R, 
\end{eqnarray}
%%%%%%%
\item consistent anomaly-matching property due to the emergence of the Nambu-Goldstone superfields together with the additional flavor-singlet composite gauge superfield.
\end{enumerate}
The low-energy symmetry is found to be:
%%%%%%%
\begin{equation}\label{Eq:FlavorSymGauged}
 U(1)^{loc}\times SO(N_f)\times U(1)_R,
\end{equation}
%%%%%%%
where $U(1)^{loc}$ is associated with the massless composite gauge superfield. 

The similar dynamical symmetry breaking is also expected for the $SO(N_c)$ theories with $N_f\geq N_c-1$, which require either a ``diquark"-type $N_c-1$ body composite of ${\hat B}^{\prime A} = \varepsilon^{AA_1\cdots A_{N_c-1}}Q^{[1\cdots}_{1\cdots}Q^{N_c-1]}_{N_c-1}$ for $N_f=N_c-1$ or a $N_c$-body composite of ${\hat B} = \varepsilon^{A_1\cdots A_{N_c}}Q^{[1\cdots}_{1\cdots}Q^{N_c]}_{N_c}$.  Equipped with these composites, we are allowed to have $\det T\neq 0$ (or equivalent configurations for $N_f \geq N_c+1$), signaling the dynamical symmetry breakdown and the Nambu-Goldstone bosons as low-energy degrees of freedom in the strongly coupled $SO(N_c)$ theories for $N_f \geq N_c-2$ \cite{Future}.
%%-------------------------------------------------
%% References
%%-------------------------------------------------

%%-------------------------------------------------
%% Tables Captions & Figure Captions
%%-------------------------------------------------
%\newpage
%\input table.tex

\begin{thebibliography}{99}
\bibliographystyle{plain}
%%-------------------------------------------------
%% References
%%-------------------------------------------------
%
\bibitem{Seiberg} 
N. Seiberg, \Journal{\PRD}{49}{6857}{1994}.

%
\bibitem{SeibergWitten} N. Seiberg and E. Witten, Nucl. Phys. {\bf B426} (1994) 19; {\bf B431} (1994) 484. 

%
\bibitem{BrokenN_2} 
R.G. Leigh and M.J. Strassler, \Journal{\NPB}{447}{95}{1995};
P.C. Argyres, M.R. Plesser and N. Seiberg, \Journal{\NPB}{471}{159}{1996};
M.J. Strassler, \Journal{\PTPSUPPL}{123}{373}{1996};
N.Evans, S.D.H. Hsu, M. Schwetz and S.B. Selipsky, \Journal{\NPSUPPL}{52A}{223}{1997};
P.C. Argyres, \Journal{\NPSUPPL}{61A}{149}{1998};
T. Hirayama, N. Maekawa and S. Sugimoto, \Journal{\PTP}{99}{843}{1998}.

%
\bibitem{EarlySeiberg}
I. Affleck, M. Dine and N. Seiberg, \Journal{\PRL}{51}{1026}{1983};
\Journal{\NPBOLD}{B241}{493}{1984}; \Journal{\NPBOLD}{B256}{557}{1985}.

%
\bibitem{CoulombPhase} 
K. Intriligator and N. Seiberg, \Journal{\NPB}{431}{551}{1994};
P.C. Argyres, M.R. Plesser and A.D. Shapere, \Journal{\PRL}{75}{1699}{1995};
A. Hanany and Y. Oz, \Journal{\NPBOLD}{B452}{283}{1995}.

%
\bibitem{Review} K. Intriligator and N. Seiberg, \Journal{\NPSUPPL}{45BC}{1}{1996};
M.E. Peskin, hep-th/9702094 in {\em Proceedings of the 1996 TASI, Boulder, Colorado, USA, 1996}; 
M. Shifman, \Journal{\PPNP}{39}{1}{1997}. 

%
\bibitem{Brane} 
See for example, A. Giveon and D. Kutasov, \Journal{\RMP}{71}{983}{1999} and other refernces therein.  

%
\bibitem{SO_IS} 
K. Intriligator and N. Seiberg, \Journal{\NPB}{444}{125}{1995}.

%
\bibitem{SO} 
N. Seiberg, \Journal{\NPB}{435}{129}{1995}; 
K. Intriligator, R. Leigh and M.J. Strassler, \Journal{\NPB}{456}{567}{1995};
C. Vafa and B. Zwiebach, \Journal{\NPB}{506}{143}{1997}; 
C. Csaki and W. Skiba, \Journal{\PLB}{415}{31}{1997}; 
T. Hirayama, N. Maekawa and S. Sugimoto, in Ref.\cite{BrokenN_2}; 
K. Fujikawa, \Journal{\PTP}{101}{161}{1999};
K. Konishi and L. Spanu, hep-th/0106175.

%
\bibitem{Conformal} 
T. Eguchi, K. Hori, K. Ito and S.-F. Yang, \Journal{\NPB}{471}{430}{1996}.

%
\bibitem{RecentSO} 
G. Carlino, K. Konishi, S.P. Kumar and H. Murayama, \Journal{\NPB}{608}{51}{2001}.

\bibitem{Alternative}  P.I. Pronin and K.V. Stepanyantz, hep-th/9902163;  K.V. Stepanyantz, hep-th/9902201;
T. Appelquist, A. Nyffeler and S.B. Selipsky, \Journal{\PLB}{425}{300}{1998}.
N. Arkani-Hamed and R. Rattazzi,  \Journal{\PLB}{454}{290}{1999};

%
\bibitem{YasueSU} 
Y. Honda and M. Yasu\`{e}, \Journal{\PTP}{101}{971}{1999}; \Journal{\PLB}{466}{244}{1999}; in {\it {Proc. the 30th Int. Conference on High Energy Physics}}, Osaka, 27 July-2 August, 2000, edited by C.S. Lim and T. Yamanaka (World Scientific, Singapore, 2001), p.1365.

%
\bibitem{YasueSO} 
M. Yasu\`{e}, \Journal{\PLB}{543}{296}{2002}.

%
\bibitem{VY} 
G. Veneziano and S. Yankielowicz, \Journal{\PLBOLD}{113B}{321}{1983};
T. Taylor, G. Veneziano and S. Yankielowicz, \Journal{\NPBOLD}{B218}{493}{1983};
G. Veneziano and S. Yankielowicz, \Journal{\NPBOLD}{B249}{593}{1985}.

%
\bibitem{MPRV} 
A. Masiero, R. Pettorino, M. Roncadelli and G. Veneziano, \Journal{\NPBOLD}{B261}{633}{1985}.

%
\bibitem{YYY} 
M. Yasu\`{e}, \Journal{\PRD}{35}{355}{1987}; \Journal{\PRD}{36}{932}{1987}; \Journal{\PTP}{78}{1437}{1987}.

%
\bibitem{tHooft} G. 't Hooft, in {\em Recent Development in Gauge Theories}, 
Proceedings of the Cargese Summer Institute, Cargese, France, 1979, edited 
by G. 't Hooft  {\em et al.}, 
NATO Advanced Study Institute Series B: Physics Vol. 59 (Plenum Press, New York, 1980).

%
\bibitem{Susskind}
S. Dimopoulos, S. Raby and L. Susskind, \Journal{\NPBOLD}{B173}{208}{1980};
T. Matsumoto, \Journal{\PLBOLD}{97B}{131}{1980}.

%
\bibitem{Future}
M. Yasu\`{e}, in preparation.
\end{thebibliography}
\end{document}